\documentclass[nato,noid,numreferences]{crckbked}
\usepackage{epsfig}
\newcommand{\eq}{\begin{equation}}
\newcommand{\eqx}{\end{equation}}
\newcommand{\eqn}{\begin{eqnarray}}
\newcommand{\eqnx}{\end{eqnarray}}

\newcommand{\nc}{$ N_{cut} $}

\begin{document}

\begin{opening}
\title{SUPERSYMMETRIC YANG-MILLS QUANTUM MECHANICS}



\author{JACEK WOSIEK}
\institute{M.Smoluchowski Institute of Physics, Jagellonian University\\
           Reymonta 4, 30-059 Krak\'{o}w, Poland}
%
%

\begin{abstract}
The new approach to quantum mechanical problems is proposed.
Quantum states are represented in an algebraic program, by lists
of variable length, while operators are well defined functions on
these lists. Complete numerical solution of a given system can
then be automatically obtained. The method is applied to
Wess-Zumino quantum mechanics and D=2 and D=4 supersymmetric
Yang-Mills quantum mechanics with the SU(2) gauge group.
Convergence with increasing size of the basis was observed in
various cases. Many old results were confirmed and some new ones,
especially for the D=4 system, are derived. Preliminary results in
higher dimensions are also presented. In particular the spectrum
of the zero-volume glueballs in  $ 4 < D < 11 $ is obtained for
the first time.
\end{abstract}

\end{opening}

\section{Introduction}

Supersymmetric quantum mechanics attracts a lot of attention for
many reasons. First, it describes well defined, finite dimensional
quantum systems which are ideal laboratory to study supersymmetry
in various disguises \cite{WI,CO,CH}. Second, the conjecture of
Banks, Fischler, Shenker and Susskind triggers additional
interest, in particular in the D=10 supersymmetric Yang-Mills
quantum mechanics (SYMQM) as a model of M-theory \cite{BFSS}. The
latter system has been studied by many authors also in lower
dimensions and for a variety of gauge groups. It is expected to
have a continuous spectrum  due to the supersymmetry driven
cancellations of zero energy fluctuations \cite{dWLN}. This is in
contrast to its non supersymmetric version, which for D=4 becomes
the zero volume limit of the well known pure Yang-Mills
gluodynamics \cite{L,LM,VABA}. Remarkably, in ten (and not less
than ten) Euclidean dimensions the continuous spectrum of the
supersymmetric system contains also a localized zero energy state.
Existence of such a threshold bound state -- the supergraviton --
is considered as a {\em sine qua non} condition for the BFSS
hypothesis \cite{POL}. Hence establishing it became an arena of
intensive and ingenious studies\cite{YI,NSM}. Interesting
numerical methods were also developed \cite{KSII}. The large N
limit of SYMQM was studied in the mean field approximation, and
interesting realizations of the black hole thermodynamics were
observed \cite{KAB3}. Simplified (i.e. quenched, D=4) systems were
also studied by lattice Monte Carlo methods for intermediate size
groups \cite{JW}. The asymptotic behavior in $N$ was observed and
a possible evidence for the two phase structure was found.

    Many interesting results have been obtained studying fully reduced (to one point) matrix model (for a review see
e.g. \cite{IKAMB}). In particular, the notorious sign problem,
which has been hindering Monte Carlo studies may have been
recently reduced \cite{AMBII}.

     In these lectures I will present a new approach to study quantum mechanical systems
\cite{JA}. The standard hamiltonian formulation of quantum mechanics will be implemented in the computer code with
the vectors in Hilbert space represented by Mathematica lists with a flexible,  dynamically varying size.
Quantum operators become well defined and simple functions on these lists. Fermionic degrees of freedom are easily
included. In practical applications one has to limit the size of the Hilbert space. However dependence on the cut-off
can be monitored, and the infinite cut-off limit extracted in many physically relevant cases.
 Of course the method becomes
computationally demanding for larger systems. Until now it has proven applicable for up to 27 degrees
of freedom.
We begin with simple two-dimensional examples: Wess-Zumino
quantum mechanics and D=2 SYM quantum mechanics. Then the new results for yet unsolved  D=4  SYMQM
will be summarized. Finally preliminary findings for higher dimensions, including D=10, will be presented.
More complete account of most of these results is in Ref\cite{JA}.

     General hamiltonian methods have been applied before to complete, space extended field
theories \cite{KS}. Recently Matsumura and collaborators
\cite{JAP} and Pinsky et al. have studied with this technique
a variety of partly reduced, supersymmetric theories in lower
dimensions (see Refs.\cite{P1} and references therein).

\section{Quantum mechanics in a PC}

 Action of any quantum mechanical observable
can be efficiently implemented in an algebraic program if we use the discrete eigen basis
 of the occupation number operator $a^{\dagger} a$
\eq
\{ |n> \}, \;\;\; |n>={1\over\sqrt{n!}}(a^\dagger)^n |0>. \label{basis}
\eqx
 The bosonic coordinate and momentum operators are
\eq
   x={1\over\sqrt{2}}(a+a^{\dagger}),\;\;\; p={1\over i \sqrt{2}}(a-a^{\dagger}), \label{XP}
\eqx
and a typical quantum observable can be
represented as the multiple actions of the basic creation and annihilation
operators\footnote{The method can be also extended
to non polynomial potentials.}.
Fermionic observables will be discussed in subsequent Sections. Generalization
to more degrees of freedom is evident and will be done there.

Any quantum state is a superposition of arbitrary number, $n_s$, of elementary states $|n>$
\eq
|st>=\Sigma_I^{n_s} a_I |n^{(I)}>, \label{stQ}
\eqx
and will be represented as a Mathematica list
\eq
st=\{n_s,\{a_1,\dots,a_{n_s}\},\{n^{(1)}\},\{n^{(2)}\},\dots,\{n^{(n_s)}\}\}, \label{stM}
\eqx
with $n_s+2$ elements. The first element specifies the number of elementary states entering the linear combination,
Eq.(\ref{stQ}), the second is the sublist supplying all complex amplitudes $a_I, I=1,\dots,n_s $,
and the remaining $n_s$ sublists give the occupation numbers of elementary, basis states.
In particular, an elementary state $|n>$ is represented by
$
 \{1,\{1\},\{n\} \}.
$

Next we implement basic operations defined in the Hilbert space: addition of two states,
multiplication by a number and the scalar product. All these can be simply programmed as definite
operations on Mathematica lists transforming them in accord
with the principles of quantum mechanics.  It is now easy to define the creation and annihilation
operators which act as a list-valued functions on above lists. This also defines the action of the position
and momentum operators according to Eq.(\ref{XP}). Then we proceed to define any quantum observables of interest:
hamiltonian, angular momentum, generators of gauge transformations, supersymmetry generators, etc.

Now our strategy is clear: given a particular system, define the list corresponding to the empty state,
then generate a finite basis of $N_{cut}$ vectors and calculate matrix representations of the hamiltonian
and other quantum operators using above rules. Given that, the complete spectrum and its various symmetry properties
is obtained by the numerical diagonalization.

Dependence of our results on the cut-off $N_{cut}$ can be monitored. In many systems studied so far
one can extract meaningful (i.e. $N_{cut}=\infty$) results
before the size of the basis becomes unmanageable.
\section{Two two-dimensional systems}
Wess-Zumino quantum mechanics (WZQM) has one complex
bosonic variable
$\phi(t)=x(t)+i y(t)\equiv x_1(t)+i x_2(t)$ and two complex Grassmann-valued fermions
$\psi_{\alpha}(t),\;\;\alpha=1,2$ \cite{SHIF}.
The hamiltonian reads (with the mass and the coupling set to 1)
\eq
H={1\over 2}(p_x^2+p_y^2)+|\phi + \phi^2|^2+\left( (1+2\phi)\psi_1\psi_2+h.c.\right),  \label{WZham}
\eqx
Bosonic creation and annihilation operators are introduced as in Eq.(\ref{XP}) for each (real) degree of freedom.
Fermionic ones -- $f_{\alpha},f_{\alpha}^{\dagger}$ -- can be chosen so that
 $\psi_{\alpha}=f_{\alpha},\;\;\psi_{\alpha}^{\dagger}=f_{\alpha}^{\dagger}$ in this case
\footnote{For the details of the  implementation of fermionic operators see \cite{JA}.}.
With these creation operators we generate
the basis which contains all (here up to two) fermionic quanta and maximum \nc\ bosonic quanta. The size of such a basis
is then $2(N_{cut}+1)(N_{cut}+2)$. With the rules of the previous section one then easily calculates matrix representation
of the hamiltonian (\ref{WZham}) in above basis. The spectrum obtained by numerical diagonalization shows rather fast
restoration of the supersymmetry: lowest state tends to zero and higher bosonic and fermionic states become degenerate
around \nc$\sim$ 10.
All this is summarized by the Witten index, Fig. 1, which is simply calculated in the energy eigen basis.

\begin{figure}[htb]
\centering
\epsfig{width=7cm,file=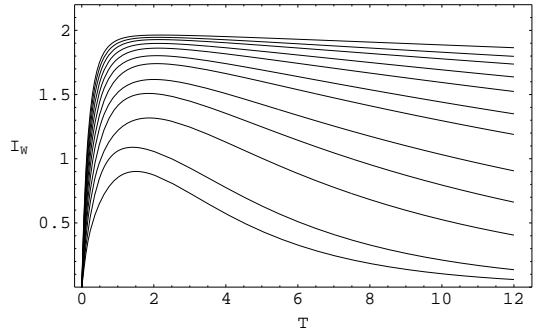}
\caption{ Witten index for Wess-Zumino quantum mechanics for
$4<$\nc$<15$. } \label{Fig1}
\end{figure}
\noindent Due to the additional symmetry in the model all states are doubly degenerate, hence the exact value $I_W(T)=2$.

\noindent {\em D=2 supersymmetric SU(2) Yang-Mills quantum mechanics.} This system has gauge invariance and
continuous spectrum,
both of which pose new challenge for our approach. It is described by three
real bosonic and three fermionic variables $x_a(t), \psi_a(t), a=1,2,3$. The hamiltonian
reads\cite{CH}
\eq
H={1\over 2} p_a p_a +i g \epsilon_{abc} \psi_a^{\dagger} x_b \psi_c={1\over 2} p_a p_a + g x_a G_a,
\label{HD2}
\eqx
with the SU(2) gauge generators $G_a=\epsilon_{abc}(x_b p_c - i \psi_b^{\dagger} \psi_c)$. To satisfy
the gauge invariance we construct the complete basis from gauge invariant states. To this end we use
four independent gauge invariant creators: $(aa)\equiv a_a^{\dagger} a_a^{\dagger},\;
 (af) \equiv a_a^{\dagger} f_a^{\dagger},\;
(aff) \equiv \epsilon_{abc}a_a^{\dagger} f_b^{\dagger}f_c^{\dagger}$, and $
 (fff) \equiv \epsilon_{abc}f_a^{\dagger} f_b^{\dagger}f_c^{\dagger}$. Since the fermionic number is conserved,
c.f. Eq.(\ref{HD2}), the last three operators (and the identity) acting on the empty state generate the four
"base" states for four sectors of the Hilbert space. All higher states are then obtained by the recursive action
of the $(aa)$ creator only. In this scheme (referred to as the 4+4+... scheme) the size of the basis is 4\nc .
Matrix representation of the hamiltonian and the spectrum are then automatically obtained as before.
The spectrum converges towards the supersymmetric multiplets and zero-energy ground state, however this time the
convergence with the cut-off is slower. This is the feature of the continuous spectrum (c.f. next Section).
Interestingly there exists a scheme of increasing the basis (2+4+4+...) where the spectrum of the
anticommutator of SUSY generators $\{Q,\bar{Q}\}$ reveals the exact supersymmetry at every finite cut-off \cite{JAP,P1,JA}.
Since the latter converges to the hamiltonian at infinite \nc\ one is free to declare it as the finite cut-off energy
operator.

Witten index vanishes identically due to the paticle-hole symmetry. However one can define the
Witten index restricted only to the two fermionic sectors ($F=0,1$ say) which are balanced by supersymmetry.
Such index does not vanish and carries nontrivial information. Very recently M. Campostrini has
calculated the index with much higher cut-off than in \cite{JA} (see Fig.2) confirming that it tends
to 1/2 at all T \footnote{I thank M. Campostrini for providing this figure prior to the publication.}.
If further established, this would provide an example of the continuous spectrum with time independent
Witten index \cite{YI}.

\begin{figure}[htb]
\centering
\epsfig{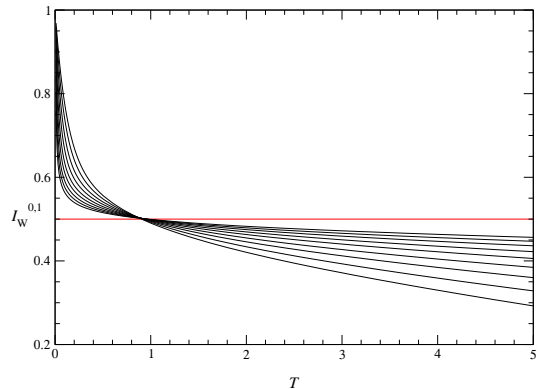}
\caption{ Restricted Witten index for D=2 SYMQM with cut-offs \nc\
$<100$.} \label{Fig2}
\end{figure}

\section{D=4 supersymmetric SU(2) Yang-Mills quantum mechanics}
Dimensionally reduced (in space) system is described by nine bosonic coordinates $ x^i_a(t) $,
$i=1,2,3; a=1,2,3$ and six independent fermionic coordinates contained in the Majorana spinor $\psi_a^{\alpha}(t)$,
$\alpha=1,...,4$.
Hamiltonian reads  \cite{HS}
\eq
H_B={1\over 2} p_a^ip_a^i+{g^2\over 4}\epsilon_{abc} \epsilon_{ade}x_b^i x_c^j x_d^i x_e^j
+ {i g \over 2} \epsilon_{abc}\psi_a^T\Gamma^k\psi_b x_c^k ,  \label{HD4}
\eqx
where $\psi^T$ is the transpose of the real Majorana spinor, and in D=4 $\Gamma$  are the Dirac $\alpha$
 matrices.  We work in the Majorana representation of Ref.\cite{IZ}.

The system has the rotational symmetry,
generated by the spin(3) angular momentum
$
J^i=\epsilon^{ijk}\left( x^j_a p^k_a-{1\over 4}\psi^T_a\Sigma^{jk}\psi_a\right),\;\;\;
\Sigma^{jk}=-{i\over 4}[\Gamma^j,\Gamma^k],
$
the gauge invariance with the generators
$
G_a=\epsilon_{abc}\left(x_b^k p_c^k-{i\over 2}\psi^T_b\psi_c \right), 
$
and $N=1$ supersymmetry with the charges
$
Q_{\alpha}=(\Gamma^k\psi_a)_{\alpha}p^k_a + i g \epsilon_{abc}(\Sigma^{jk}\psi_a)_{\alpha}x^j_b x^k_c.
$
Again, expressing bosonic observables in terms of creation and annihilation operators is evident. For fermions
we use
\eq
\psi_a={1+i\over 2\sqrt{2}} \left( \begin{array}{c}
                            -   f_a^{1} - i f_a^{2} + i f_a^{1\dagger} +   f_a^{2\dagger} \\
                            + i f_a^{1} -   f_a^{2} -   f_a^{1\dagger} + i f_a^{2\dagger} \\
                            -   f_a^{1} + i f_a^{2} + i f_a^{1\dagger} -   f_a^{2\dagger}  \\
                            -i  f_a^{1} -   f_a^{2} +   f_a^{1\dagger} + i f_a^{2\dagger}  \\
                                    \end{array} \right),
\eqx
which gives the quantum hermitean Majorana spinor satisfying
$\{\psi_a^{\alpha},\psi_b^{\beta}\}=\delta_{ab}^{\alpha\beta}$.
Fermionic operators are implemented
with the aid of the Jordan-Wigner construction using the lexicographic ordering of the double
index $A=(\sigma,a)$. It turns out that the fermionic number $F=f_a^{\sigma\dagger}f_a^{\sigma}$ is conserved.
Consequently the hamiltonian can be diagonalized independently in all seven fermionic sectors,
which is done\footnote{Construction of the basis is similar to the D=2 SYMQM but more involved and will
not be discussed here. Full details are in Ref.\cite{JA}.}
analogously to the previous cases.

\begin{figure}[htb]
\centering
 \epsfig{width=10cm,file=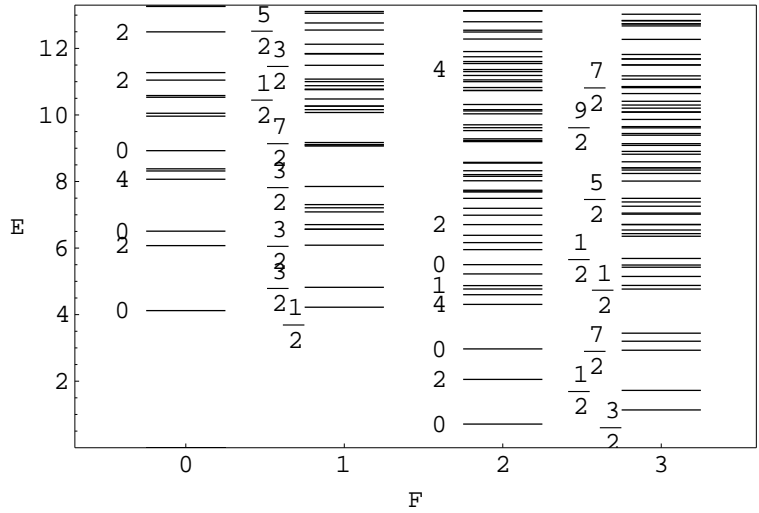}
\caption{Spectrum of D=4 SYMQM.
}
\label{Fig3}
\end{figure}

The spectrum of the theory is shown in Fig.\ref{Fig3}. A sample of
states is labeled with their angular momenta $j$, which are simply
assigned using the matrix representation of the angular momentum.
It is important that our cut-off preserves the SO(3) symmetry. In
particular the states have appropriate degeneracies for each value
of $j$. For current $N_{cut}$ the spectrum extends to $E_{max}\sim
35$.

Sectors with $F=6,5,4$ can be obtained from those with $F=0,1,2$
by the particle-hole transformation. Since our cut-off respects
this symmetry we indeed observe, exact repetition of the
eigenvalues in corresponding subspaces. Therefore only first four
sectors are displayed in Fig.\ref{Fig3}. Evidently the spectrum of
D=4 theory is very rich. We begin its discussion with the
correspondence to the pure Yang-Mills quantum mechanics.

The fermionic part of the hamiltonian,
Eq.(\ref{HD4}) vanishes on the $F=0$ sector. Therefore our effective hamiltonian in this sector is nothing but
that of the zero volume pure Yang-Mills theory considered by L\"{u}scher \cite{L}. Indeed we observe rather satisfactory
convergence of the $F=0$ eigen energies to the results of L\"{u}scher and M\"{u}nster \cite{LM} , c.f. Fig.4.

\begin{figure}[htb]
\centering
 \epsfig{width=7 cm,file=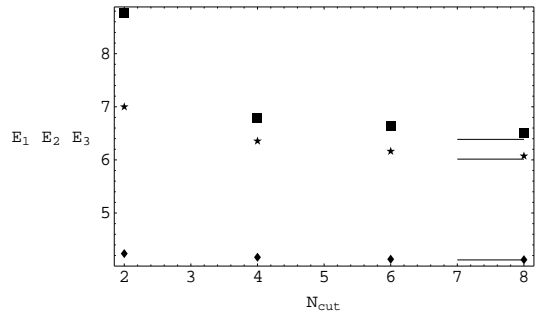}
\caption{ Lower energy levels in $F=0$ sector compared with the
zero volume glueballs of Ref.\cite{LM}} \label{Fig4}
\end{figure}

Another comparison was done together with van Baal who studied $F=2$ sector in a more complex,
non compact, supersymmetric model of the same family \cite{vBN}. At the time of writing \cite{JA} our results
could not be directly compared, due to the specific boundary conditions required in \cite{vBN}. However after
this conference van Baal has adapted his programs so that they can be applied also to our situation. We have found
complete agreement to the full Mathematica precision in both $F=0$ and $F=2$ sectors\footnote{I thank P. van Baal for
providing his results and programs.} .

\begin{figure}[htb]
\centering
 \epsfig{width=10 cm,file=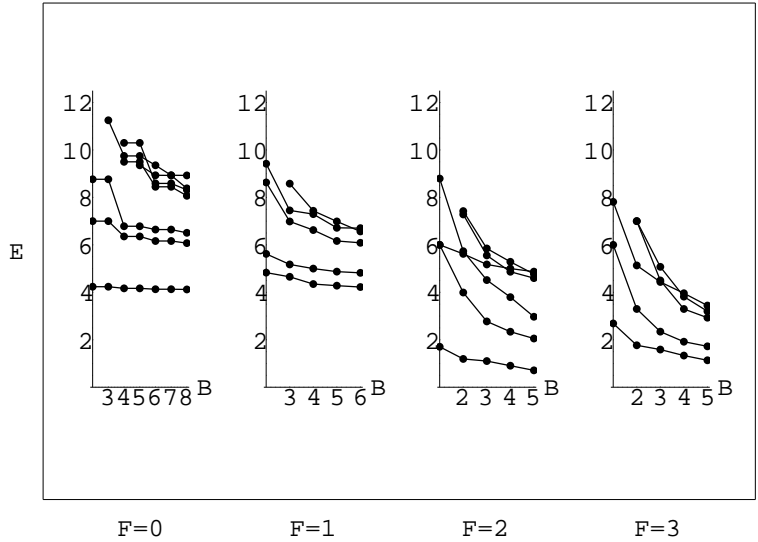}
 \caption{Cut-off dependence of
the lower levels of the D=4 SYMQM in four independent fermionic
channels (\nc=B). } \label{Fig5}
\end{figure}

Next we discuss the convergence with the cut-off (c.f. Fig.5). The first striking property
is that the \nc\  dependence is different in various fermionic sectors.
For $F=0,1,5,6$ the dependence is weak, hence the lowest states have
already converged (within some reasonable precision). For $F=2,3,4$ however, the convergence is slower.
We interpret this as an indication that
the spectrum is discrete in the $F=0,1,5,6$ sectors while it is continuous in the "fermion/hole-rich" $F=2,3,4$ environments.
This is based on our finding of an interesting and useful correlation between the localizability of a state
and the rate of the convergence with the cut-off \nc . It turns out that the spectrum of the momentum operator
in the cut-off Hilbert space is exactly given by the zeroes of the Hermitte polynomials \cite{TW}. Consequently the
convergence of the free spectrum is slow $\sim O(1/N_{cut} )$. On the contrary, for the localized (i.e. bound) states
the simple argument suggests that the rate of convergence is given by the large x asymptotics of the wave function,
hence it is fast (e.g. exponential). This different nature of the spectra in different fermionic sectors provide
the reasonable extension of the result of Ref.\cite{dWLN} which was mentioned earlier.
Since the fermionic modes are crucial to generate
the continuous spectrum, it is natural that it does not show up in the sectors where they do not exist at all,
or are largely freezed out by Pauli blocking.
On the other hand not all states in $F=2,3,4$ sectors have to belong to the continuum.
Supersymmetry together with the discrete spectrum in $F=0,1$ channels implies existence of the normalizable
states among the continuum of $F=2,3,4$ states. As one explicit example we give
 the energy of the $F=2, j=1$ state, shown in Fig. 5 (flatter curve beginning at E=6). It has definitely weaker
dependence on \nc\ than the others.
This indicates that the lowest $|2_F,1_j>$ state is localized. This  may be a precursor
of the more complex phenomenon expected in the D=10 theory. There, the zero energy localized bound state
of $D_0$ branes  should exist at the threshold of the continuous spectrum. Present example suggests that one way to
distinguish such a state from the continuum may be by the different \nc\ dependence. Finally our results show
that the supersymmetric vacuum cannot be in the $F=0$ sector. Instead the best candidates are the lowest states
in the $F=2$ and $4$ sectors.

\begin{figure}[htb]
\centering
 \epsfig{width=8cm,file=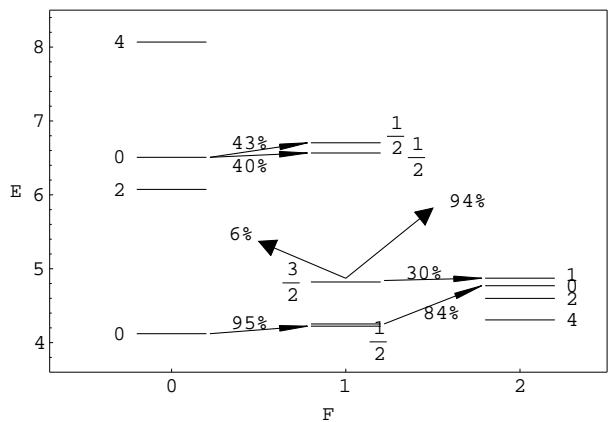}
\caption{ Supersymmeric images of a sample of eigenstates.}
\label{Fig6}
\end{figure}

{\em SUSY on the level of states.}
Evidently identifying SUSY multiplets in Fig.3 looks like a nontrivial task. However present approach
allows to monitor directly the action of the SUSY generators on the eigenstates of the hamiltonian.
As any other observables they can be implemented in our "quantum algebra" described in the second Section.
Alternatively their matrix representations are easily obtained once bases are constructed in each sector.
Then the supersymmetric image of an eigenstate from one sector can be decomposed into states from adjacent sectors
(the charges change fermionic number by $\pm 1$). This allows identification of the approximate supersymmetry partners.
With the current scheme supersymmetry is restored only at infinite \nc , hence the energies of the SUSY partners
are not exactly identical. Neither the SUSY images of eigenstates are 100\% concentrated on  single eigenstates.
This situation is quantitatively summarized in Fig.6. The lower multiplets can be already identified, while
 for higher states, larger cut-offs are required for better assignment of SUSY partners, which is not surprising.

\noindent {\em Witten index.} Figure 7 shows the (Euclidean) time dependence of the Witten index
 calculated from our spectrum with up to five bosonic quanta.

\begin{figure}[htb]
\centering
 \epsfig{width=8cm,file=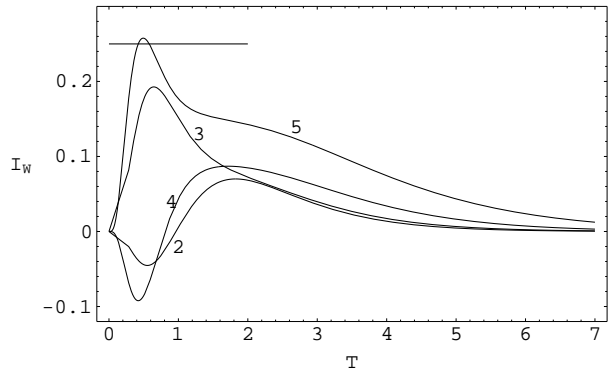}
\caption{ Witten index of the D=4, supersymmetric Yang-Mills quantum mechanics for the number of bosonic quanta
bounded by $N_{cut}=2,...,5$. The bulk value 1/4 is also shown. }
\label{Fig7}
\end{figure}

Clearly we are far from the convergence, but some interesting signatures are already there.
First, one can see how the discontinuity at $T=0$ emerges even at this early stage. Second, the exponential
fall-off common to all curves at large $( T > 5 )$ times is evident. This is where the lowest state dominates.
Since at this \nc\ its energy is not zero, the fall-off is exponential. Finally, and most interestingly,
we observe the flattening shoulder appearing at $T\sim 2-3$. This is the signal of the supersymmetric cancellations
which occur on the average, even though the exact SUSY correspondence between individual states does not yet show up.
The behavior with \nc\ is not inconsistent with the exact bulk value 1/4  obtained from the
nonabelian integrals \cite{YI,NSM}. Clearly higher \nc\ are needed, and, what is important, they are within
 the range of present computers.
\section{From D=4 to D=10 space-time dimensions}
To bridge the gap to ten dimensions we study first the simpler, $F=0$ case in all
space dimensions $ 2 < d=D-1 < 10 $. Thereby results presented here extend for the first time
the spectrum  of the zero volume glueballs to higher dimensions. To this end we use the hamiltonian
in Eq.(\ref{HD4}) without the fermionic part, drop all the fermionic occupation numbers in elementary states,
Eq.(\ref{stM}), and appropriately extend the ranges of spatial indices.
Therefore generalization to higher dimension is trivial in our approach,
however it puts considerably heavier load on the computer.
Hence our results are rather preliminary
and will be updated soon. Figure 7 shows the energies of the first three levels as a function of d.

\begin{figure}[htb]
\centering
 \epsfig{width=8cm,file=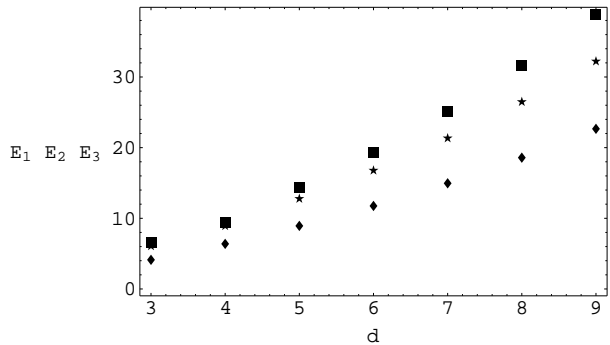}
\caption{ Three lowest glueball states in various dimensions. }
\label{Fig8}
\end{figure}
The higher dimensional glueballs are there as expected. The
ordering of the SO(d) multiplets remains the same as in d=3:
scalar, two dimensional symmetric tensor with dimension
g=d(d+1)/2-1, and scalar again. Our cut-off respects rotational
symmetry for all d, hence we observe these degeneracies exactly in
the spectra. The largest uncertainty (i.e. that of the d=9 points)
is around 25\% at present. Given above range of d one may attempt
to confront these results with the large d expansions. The
eye-ball fit suggest the power dependence  $ E \sim d^{\sim 1.5}
$.

We conclude that calculations in higher dimensions including $d=9$ are perfectly realistic (for SU(2) gauge group)
even if much more time consuming. Including fermions will increase this demand by approximately factor of 16.

\section{Summary}

The new approach to solve quantum mechanical systems on a computer is proposed. The Hilbert space of quantum states
is implemented algebraically in a computer code with states represented
as flexible Mathematica lists. Any quantum observable is then defined as the list valued function on these lists.
This allows for the automatic calculation of matrix representations and the spectrum. A range of progressively
more complicated (also supersymmetric) systems from two to ten space-time dimensions was studied and many new
results were obtained. The method is practically applicable to the D=10 supersymmetric SU(2) Yang-Mills quantum mechanics,
which is our main goal in the future. Extension to few higher gauge groups is conceivable but large N require
dedicated and new modifications.

\noindent {\em Acknowledgments.} I thank Jeff Greensite and Stefan
Olejnik for providing the opportunity to present this work, for
financial support, and for the warm atmosphere enjoyed during the
Conference. I would like to thank C. M. Bender for an instructive
discussion which inspired this approach. I also thank P. van Baal,
P. Breitenlohner and H. Saller
 for intensive discussions. This work is supported by the
Polish Committee for Scientific Research under the grants no. PB 2P03B01917 and PB 2P03B09622.

\end{document}